# Optical conductivity and optical effective mass in a high-mobility organic semiconductor: Implications for the nature of charge transport

Yuan Li,[1] Yuanping Yi,[1,2] Veaceslav Coropceanu,[1,*] and Jean-Luc Brédas[1,†]

[1]*School of Chemistry and Biochemistry & Center for Organic Photonics and Electronics, Georgia Institute of Technology, 901 Atlantic Drive NW, Atlanta, Georgia 30332-0400, USA*

[2]*Key Laboratory of Organic Solids, Beijing National Laboratory for Molecular Sciences (BNLMS), Institute of Chemistry, Chinese Academy of Sciences, Beijing 100190, China*

**ABSTRACT**

We present a multiscale modeling of the infrared optical properties of the rubrene crystal. The results are in very good agreement with the experimental data that point to nonmonotonic features in the optical conductivity spectrum and small optical effective masses. We find that, in the static-disorder approximation, the nonlocal electron-phonon interactions stemming from low-frequency lattice vibrations can decrease the optical effective masses and lead to lighter quasiparticles. On the other hand, the charge-transport and infrared optical properties of the rubrene crystal at room temperature are demonstrated to be governed by localized carriers driven by inherent thermal disorders. Our findings underline that the presence of apparently light carriers in high-mobility organic semiconductors does not necessarily imply band-like transport.

**PACS numbers:** 72.80.Le, 72.10.−d, 71.38.−k

**Keywords:** Organic Semiconductor, Optical Conductivity, Effective Mass, Charge Transport, Electron-Phonon Coupling

---

[*] coropceanu@gatech.edu
[†] jean-luc.bredas@chemistry.gatech.edu

# I. INTRODUCTION

Efficient organic electronic devices, such as organic field-effect transistors (OFETs), solar cells, or light-emitting diodes, require high charge-carrier mobilities in the active organic layers.[1] An increasing number of molecular organic semiconductors have been recently reported with carrier mobilities as high as tens of cm$^2$/Vs near room temperature.[2–5] These mobilities generally display a power-law temperature dependence, i.e., $\mu \propto T^{-n}$ ($n > 0$). Well-defined electronic bands were also observed by angle-resolved photoemission spectroscopy (ARPES).[6–9] The charge-transport properties of such high-mobility organic semiconductors were usually interpreted in the context of a band-like mechanism, a scenario similar to that of their inorganic counterparts in terms of the motion of wave packets under successive scattering events. However, while the electron-phonon (e-ph) couplings in these high-mobility materials are not strong enough to lead to substantial polaronic effects,[10–14] thermal disorders can readily appear as a result of the weak Van der Waals interactions among molecules and the interplay between crystal packing and transfer integrals.[15,16] Large thermal disorders near room temperature in organic crystals can significantly localize charge carriers and thereby cause a breakdown of the band-like transport.[17–23] A puzzling aspect is that the carrier effective masses at room temperature[6,7] generally present values very close to or even smaller than those derived from band-structure calculations in the limit of zero temperature;[13,24] such features would again suggest that a band-like mechanism is operational in these systems. Therefore, it is essential to explore the origin of the small effective masses in organic crystals and clarify whether the presence of light carriers, as required for achieving high mobility in the framework of band theory, does necessarily point to band-like transport.

As a powerful tool to probe carrier dynamics, infrared spectroscopy was recently employed to measure the optical conductivity and effective masses in OFET configurations based on the high-mobility rubrene crystal.[25,26] In contrast to the typical Drude response with a monotonic roll-off starting from zero energy,[27] the optical conductivity spectra in the rubrene crystal feature a broad peak centered at a finite energy, reminiscent of disordered systems near the metal-insulator transition.[28] However, a signature of light quasiparticles was also observed, with effective masses as small as those from band-structure calculations;[13,24] such results suggest the involvement of band-like states pertaining to ordered systems. While these contrasting observations pose a challenge, a detailed theoretical investigation, to the best of our knowledge, has not yet been performed.

In this work, we address the issues raised above via a multiscale modeling protocol to simulate the infrared optical properties of high-mobility organic semiconductors. When applied to the rubrene crystal, the experimental nonmonotonic optical conductivity spectra and small effective masses are very well reproduced and understood via a model that adequately accounts for the thermal disorders in the system. This leads to new insights into the infrared properties and their connection to the electronic and transport properties in organic crystals.

## II. METHODOLOGY

We consider the electronic properties along the highest-mobility direction of an organic crystal. Accordingly, the following one-dimensional tight-binding Hamiltonian is used:

$$H = \sum_n \varepsilon_n a_n^+ a_n + \sum_n t_{n,n+1}\left(a_n^+ a_{n+1} + \text{H.c.}\right) + \sum_{nj} \frac{\hbar \omega_{\text{ph}j}}{2}\left(u_{nj}^2 + p_{nj}^2\right). \quad (1)$$

Here, $a_n^+$ ($a_n$) denotes the electronic creation (annihilation) operator at site $n$; $\varepsilon_n$, the molecular on-site energy; and $t_{n,n+1}$, the nearest-neighbor transfer integral. $\omega_{\text{ph}j}$, $u_{nj}$ and $p_{nj}$ represent the frequency, dimensionless coordinate and momentum of the $j$th vibration mode, respectively. $\hbar$ denotes the Planck's constant. Given that the thermal disorders in prototypical organic crystals mainly come from the nonlocal e-ph couplings,[10,11,16,29] i.e., the modulation of transfer integrals by lattice vibrations,[30] we focus here on the nonlocal mechanism and set $\varepsilon_n = 0$. In the linear coupling approximation, the transfer integral incorporating the nonlocal e-ph couplings reads

$$t_{n,n+1} = t^{(0)} + \sum_j \left[ \upsilon_{sj}\left(u_{nj} + u_{n+1,j}\right) + \upsilon_{aj}\left(u_{nj} - u_{n+1,j}\right) \right], \qquad (2)$$

where $t^{(0)}$ represents the transfer integral at the equilibrium geometry, and the terms associated with $\upsilon_{sj}$ and $\upsilon_{aj}$ denote the symmetric and antisymmetric nonlocal e-ph couplings,[23,31] respectively. The symmetric and antisymmetric mechanisms, as will be shown, represent important components that both have to be incorporated in the model. The magnitude of thermal disorders is defined by the variance of the transfer integrals $\Delta^2 = \langle t_{n,n+1}^2 \rangle - \langle t_{n,n+1} \rangle^2$, where $\langle \cdots \rangle$ denotes the thermal average over the lattice phonons.[10] In the classical limit, we obtain $\Delta^2 = 2Lk_BT$, where $T$ denotes the temperature, $k_B$ the Boltzmann constant, $L = L_s + L_a$ the nonlocal relaxation energy, and $L_{s[a]} = \sum_j \upsilon_{s[a]j}^2 / \hbar \omega_{phj}$ the contribution from the symmetric [antisymmetric] coupling. To evaluate the e-ph coupling symmetry, we define parameter $c = L_s/L$, where $1 \geq c \geq 0$ and $c = 1$ and $c = 0$ denote the symmetric and antisymmetric couplings, respectively. Importantly, as the symmetry effects essentially represent the correlation among transfer integrals,[23] we obtain that:

$$c = \frac{\langle \delta t_{n-1,n} \delta t_{n,n+1} \rangle}{\Delta^2} + \frac{1}{2}, \qquad (3)$$

where $\delta t_{mn} = t_{mn} - \langle t_{mn} \rangle$ and the associated term denotes the correlation function between adjacent transfer integrals.

The model parameters can be estimated at multiple scales by quantum-chemical calculations coupled with molecular dynamics (MD) simulations. The parameters computed specifically for the rubrene crystal along the $b$ (stacking) direction, as shown in the inset of Fig. 1, are collected in Table I. For the sake of simplicity, only one effective vibration mode with frequency $\omega_{ph}$ was considered. On the basis of the experimental crystal structure,[32] the transfer integral $t^{(0)}$ was calculated at the density functional theory (DFT) B3LYP/6-31G(d,p) level in combination with a basis set orthogonalization procedure.[33] $\Delta^2$ and $c$ were computed by using DFT as well as a semiempirical Hartree-Fock method at the intermediate neglect of differential overlap (INDO) level coupled with MD simulations, which can account for all the low-frequency vibrations that have major contributions to the nonlocal e-ph couplings.[34] In practice, based on the MD geometries, we first computed the ratio $\Delta/t^{(0)}$ at a given temperature $T$ at the INDO level; making the reasonable assumption that the value of $\Delta/t^{(0)}$ calculated with INDO applies to DFT as well, we rescaled $\Delta$ accordingly with respect to the DFT value of $t^{(0)}$ and then obtained the nonlocal relaxation energy following $L = \Delta^2/2k_B T$. Similarly, we first computed the correlation function $\langle \delta t_{n-1,n} \delta t_{n,n+1} \rangle / \Delta^2$ and then obtained the e-ph coupling symmetry parameter $c$ according to Eq. (3). The MD simulations were performed with the MM3 force field by using the TINKER package.[35] We underline that the method presented here can be readily extended to any other organic crystal in which thermal disorders play a significant role.

### III. OPTICAL CONDUCTIVITY AND OPTICAL EFFECTIVE MASS

Based on the parameters estimated above, we first calculated the optical conductivity spectra of the rubrene crystal. According to linear response theory,[36] the real part of the optical conductivity (for holes) in the spectral representation can be calculated from:

$$\sigma(\omega) = \left\langle \frac{\pi(1-e^{-\beta\hbar\omega})}{Nb\hbar\omega} \sum_{jl} f_j(1-f_l) |\langle l|J|j \rangle|^2 \delta(\hbar\omega + E_l - E_j) \right\rangle_{\{u_n\}} . \quad (4)$$

Here, $\omega$ represents the photon frequency; $\beta = 1/k_B T$, the thermodynamic beta; $b$, the lattice constant; $N$, the number of sites; $J = ieb \sum_n t_{n,n+1} \left( a_n^+ a_{n+1} - \text{H.c.} \right)$, the current density operator; $e$, the elementary charge; $f_j = \frac{1}{e^{\beta(E_f - E_j)} + 1}$, the Fermi function; $E_f$, the Fermi level; $E_j$ and $|j\rangle$, the $j$th eigenenergy and eigenfunction of the electronic subsystem, respectively. $\langle \cdots \rangle_{\{u_n\}}$ denotes the thermal average over the lattice configurations $\{u_n\}$ in the static-disorder limit (setting $\{p_n\} = 0$). In order to better identify the impact of thermal disorder on the issues of interest, the electron self-trapping caused by e-ph couplings was not considered in the calculations and we performed the thermal average over a non-distorted lattice. We note that such approximations have been shown to be adequate in treating the nonlocal e-ph couplings stemming from low-frequency lattice vibrations in high-mobility organic crystals.[12,20,23] In practice, the Dirac $\delta$-function in Eq. (4) was replaced by a Lorentzian function with a width of 7.5 cm$^{-1}$, and a lattice of 256 sites was used throughout the calculations. The results in the limit of low carrier concentration (one-hole case) are presented in Fig. 1. The calculated optical conductivity spectra near room temperature generally present a broad peak centered at a finite energy, which reproduces the experimental observations.[25,26] The peak attenuates and disperses with increasing temperature. At 300 K, the optical conductivity spectrum peaks around 340 cm$^{-1}$, which agrees well with the experimental data of about 400−500 cm$^{-1}$. The minor difference is likely related to the presence of defects and impurities in actual samples and/or the impact of local e-ph couplings[37] that are not incorporated in the model.

The effective mass can be generally obtained by applying the Drude model:[27]

$$\sigma(\omega) = \frac{n_0 e^2 \tau}{m^*} \frac{1}{1 + \omega^2 \tau^2}, \tag{5}$$

where $n_0$ represents the carrier number density and $\tau$ the relaxation time. By integrating Eq. (5) over the photon frequency, we obtain that:

$$\frac{1}{m^*} = \frac{2}{\pi n_0 e^2} \int_0^\infty \sigma(\omega) d\omega, \qquad (6)$$

where $m^*$, usually referred to as the *optical effective mass*, is directly accessible through the optical conductivity probe.[26] Alternatively, the effective mass can also be evaluated through the standard definition:

$$\frac{1}{m_{\text{eff}}} = \frac{1}{\hbar^2}\left|\frac{\partial^2 E_k}{\partial k^2}\right|, \qquad (7)$$

where $E_k$ denotes the electronic band dispersion. To distinguish it from the optical effective mass, we refer to $m_{\text{eff}}$ in Eq. (7) as the *band effective mass*. By taking $b = 7.2\,\text{Å}$ in the rubrene crystal, the band effective mass at the top of the valence band ($k = 0$) when neglecting e-ph coupling ($L = 0$) is estimated to be $m_{\text{eff}} = 0.77 m_e$ (with $m_e$ the electron rest mass), in good agreement with the results of band-structure calculations at the DFT level.[13] We also computed the optical effective mass following Eq. (6); the results as a function of e-ph coupling are presented in Fig. 2(a). Small $m^*$ values comparable to $m_{\text{eff}}$ are obtained, in accordance with the experimental observations.[26] Interestingly, however, *the optical effective mass monotonically decreases with increasing e-ph coupling strength*. For instance, at 300 K, $m^*$ for $L = 21.4\,\text{meV}$ is about 20% smaller than that in the absence of e-ph coupling. To interpret this finding, we performed perturbative calculations in the spirit of Boltzmann theory and, to first-order approximation, obtain that:

$$\frac{1}{m^*} = \frac{1}{m^*_{(0)}} + \frac{1}{m^*_{(1)}}$$

$$\frac{1}{m^*_{(0)}} = 2b^2 \sum_{jl} \frac{e^{+\beta E_j}}{Z_e} \left|M^{(0)}_{lj}\right|^2 F(0) \qquad (8)$$

$$\frac{1}{m^*_{(1)}} = \frac{b^2}{(1-e^{-\beta})} \sum_{jl} \frac{e^{+\beta E_j}}{Z_e} \left|M^{(1)}_{lj}\right|^2 \left[F(-1) + F(1)e^{-\beta}\right]$$

Here, $E_j = \sum_k \phi_{jk}^2 E_k$ with $\phi_{jk}$ the unperturbed electronic wavefunction and $E_k = 2t^{(0)} \cos(kb)$; $Z_e = \sum_j e^{+\beta E_j}$ ; $F(x) = \frac{1-\exp\{-\beta[x+\sum_k(\phi_{jk}^2-\phi_{lk}^2)E_k]\}}{x+\sum_k(\phi_{jk}^2-\phi_{lk}^2)E_k}$ ; $M^{(0)}_{lj} = \sum_k \phi^*_{lk}\phi_{jk}t_k$ with $t_k = i2t^{(0)} \sin(kb)$; and $M^{(1)}_{lj} = \sum_{kk'} \phi^*_{lk'}\phi_{jk}\upsilon_{k'k}$ with $\upsilon_{k'k} = 2\sqrt{L/N}\left\{\sqrt{1-c}[\cos(k'b)-\cos(kb)] + i\sqrt{c}[\sin(k'b)+\sin(kb)]\right\}$.

We note that $m^*_{(0)}$ and $m^*_{(1)}$ represent the zeroth-order optical effective mass and the first-order correction, respectively. As $m^*_{(0)}$ is independent of $L$ and $m^*_{(1)} \propto 1/L$, the total optical effective mass is inversely proportional to the e-ph coupling strength, which is in line with the numerical results. Also, it is seen from Fig. 2(a) that the optical effective masses present opposite temperature dependences for weak and strong e-ph couplings, with a crossover taking place at moderate strength of the couplings. This can be rationalized according to Eq. (8) since $m^*_{(0)} \propto T$ (when $T \to \infty$) and $m^*_{(1)} \propto e^{1/k_B T}$ (when $T \to 0$), with these parameters predominating in the weak and strong coupling regimes, respectively, as shown in Fig. 3. We underline that, as Eq. (8) was derived according to Boltzmann theory in which phonons are treated fully quantum-mechanically,[38] the agreement between the analytical and numerical calculations indicates that our results are not subject to the static-disorder approximation taken in Eq. (4).

While small optical effective masses can indeed be explicitly obtained, this does not guarantee the validity of a band-like transport mechanism in the rubrene crystal. In Fig. 2(b), we display the mean free path of the charge carrier and the energy of the optical conductivity peak as a function of e-ph coupling at 300 K. The carrier mean free path was calculated by $\langle \xi \rangle = Z^{-1} \sum_k e^{+\beta E_k} \xi_k$, where $Z = \sum_k e^{+\beta E_k}$ and $\xi_k = \tau_k |v_k|$ represents the mean free path for state $k$ with $v_k = \partial E_k / \partial k$ the corresponding electronic velocity and $\tau_k$ the relaxation time.[38] As seen from Fig. 2(b), the transport properties reach the Ioffe-Regel-Mott (IRM) limit,[39,40] i.e., $\langle \xi \rangle = b$, at even moderate strength of the coupling ($L = 11.5$ meV), indicating that band-like transport breaks down in the rubrene crystal (where $L = 21.4$ meV) regardless of the presence of carriers with very small effective masses. This paradoxical result essentially stems from the dual roles played by the thermal disorders in organic crystals. On the one hand, the charge carriers in real space are strongly scattered and thereby localized by the thermal disorders near room temperature, leading to the breakdown of band-like transport.[17–23] On the other hand, the quasiparticle states in reciprocal space are substantially broadened by the thermal disorders, giving rise to an overall *broadening*, rather than narrowing, of the electronic bandwidths;[12,20,23] in such an instance, the "effective" transfer integral $t^*$ increases as a result of thermal disorder and, accordingly, the band effective mass decreases through $1/m_{\text{eff}}^* = 2\hbar^{-2} b^2 t^*$, which can be obtained from Eq. (7) by assuming $E_k = 2t^* \cos(kb)$.

It is important to note that the above result is in striking contrast to the conventional wisdom that the band effective mass is always enhanced by e-ph couplings, *i.e.*, $m_{\text{eff}}^* \approx m_{\text{eff}}(1+\lambda)$, where $m_{\text{eff}}^*$ represents the band effective mass in the presence of e-ph coupling and $\lambda$ denotes the mass enhancement factor.[36] This is due to the fact that the e-ph coupling in the present work is treated in the static approximation and thus neglects the dynamical effect, which is responsible for band effective mass enhancement and electronic bandwidth narrowing according to polaron theories.[41,42] It should be noted that the results of electronic structure calculations for high-mobility organic crystals indicate that the nonlocal e-ph coupling is weak and related mainly to low-energy phonons, justifying the use of the static approximation.[10-14] To gain a better understanding, further investigations based on a dynamical treatment of the nonlocal e-ph couplings are clearly needed. We note that previous studies based on the variational principle have demonstrated that the nonlocal e-ph couplings can renormalize the electronic band by introducing new minima and ultimately lead to its broadening,[43,44] which, in line with our analysis above, could indeed result in smaller band effective masses. However, the band effective mass should not be confused with the optical effective mass since the latter, which is obtained from the *f*-sum rule [Eq. (6)] rather than from the band structure, can be expected to decrease with increasing nonlocal e-ph couplings even when the band-narrowing effect is operational in the system.[41]

We also note that a thermal narrowing of the electronic bandwidths has been shown via ARPES in the pentacene crystal;[7,9] however, it has been recently demonstrated that this result comes from the impact of crystal thermal expansion rather than a renormalization of e-ph couplings.[13] In the inset of Fig. 2(a), we contrast the temperature dependence of the optical effective mass in the rubrene crystal with and without considering the impact of crystal thermal expansion. It is seen that $m^*$ is instead enhanced with increasing temperature when the crystal thermal expansion is taken into account. As the optical effective mass in both cases presents weak temperature dependence, it is expected that our results can be experimentally confirmed via temperature-dependent probes.

## IV. ORIGIN OF OPTICALCONDUCTIVITY PEAK

While a finite-energy optical conductivity peak is characteristic of disordered systems,[21,28] its origin with respect to the corresponding optical absorption in organic crystals remains unclear. In the following, we address this issue by examining the impacts of the e-ph coupling symmetry and Fermi energy.

We first evaluate the optical conductivity properties as a function of the e-ph coupling symmetry, which has a significant impact on the electronic and transport properties of organic crystals.[23,38] As seen from Fig. 4(a), for an e-ph coupling representative of the rubrene crystal, the optical conductivity spectra are markedly different in lineshape as a function of the e-ph coupling symmetry. A remarkable blue-shift and a broadening of the optical conductivity peak are observed when going from antisymmetric ($c=0$) to symmetric ($c=1$) couplings. Interestingly, in the case of antisymmetric coupling, in addition to the main peak ($\hbar\omega_1$) at low energy, two sub-peaks emerge at higher energies near $2t^{(0)}$ ($\hbar\omega_2$) and $4t^{(0)}$ ($\hbar\omega_3$). These results clearly indicate that the e-ph coupling symmetry is a major ingredient to be taken into account in the simulations.

In order to describe the associated optical processes, we introduce the quantity $\Xi(E_{ini}, E_{fin})$ that is connected to the optical conductivity by:

$$\sigma(\omega) = \int \Xi(E, E - \hbar\omega) dE . \tag{9}$$

Here, $E_{ini}$ and $E_{fin} = E_{ini} - \hbar\omega$ denote the energies of the initial and final states relevant upon absorption of a photon with energy $\hbar\omega$; $\Xi(E_{ini}, E_{fin})$ thus represents the "energy-resolved" optical conductivity weighted by the electronic density of states (DOS):

$$\Xi(E_{ini}, E_{fin}) = \left\langle \sum_{jl} \Xi_{jl} \delta(E_{ini} - E_j) \delta(E_{fin} - E_l) \right\rangle_{\{u_n\}} , \tag{10}$$

where $\Xi_{jl}$ is expressed as:

$$\Xi_{jl} = \frac{\pi}{Nb} \frac{\left(1 - e^{-\beta(E_j - E_l)}\right)}{E_j - E_l} f_j (1 - f_l) |\langle l|J|j\rangle|^2 . \tag{11}$$

Here, $E_j - E_l \geq 0$ for absorption by holes. The results for the symmetric and antisymmetric couplings are presented in Figs. 4(b) and 4(c), respectively. In the case of symmetric coupling, the optical absorptions take place around the band tail ($E > 2t^{(0)}$), which is the region where the electronic states are mainly occupied at thermal equilibrium. The contributions to the optical conductivity peak (as indicated by the white line in the figure) represent the absorption that manifests itself most strongly. In the case of antisymmetric coupling, a similar behavior is observed for the main peak of the optical conductivity spectrum; for the other two sub-peaks, however, it turns out that the absorptions take place between states near the band edge and the band center (for $\hbar\omega_2$) or between those near the two band edges (for $\hbar\omega_3$), which corresponds to the energy regimes where the DOS features singularities. We note that these singularities, *i.e.*, the Van Hove singularities at the band edges and the anti-localization peak at the band center, are characteristics of the one-dimensional model we consider.[23] In materials of a three-dimensional nature, these singularities are in general absent in the DOS;[13] thus, the two sub-peaks in the optical conductivity spectra should not be expected in actual experiments.[25,26] However, this result is interesting in the sense that it strongly suggests that attention should be paid to the DOS singularities when analyzing peculiar structures in optical conductivity spectra.

We now turn to a discussion of the impact of the system Fermi energy. As in OFETs the carrier concentration and occupation within the electronic spectrum are tunable via gate voltage, it is possible to access the states beyond the band tail that were expected to be responsible for band-like transport in organic crystals.[20] To model this situation, we performed Fermi-level dependent calculations where we ignore the interactions between carriers. As shown in Fig. 5(a), the energy of the optical conductivity peak is significantly modified by the position of the Fermi level, and distinct behaviors are obtained as a function of the e-ph coupling symmetry. For instance, when the Fermi level is close to the band center ($E = 0$), the optical conductivity peak is larger in energy for the cases with stronger contributions from antisymmetric coupling, while it is the opposite when the Fermi level is far away from the energy band center. This is related to the fact that the nonlocal e-ph coupling manifests most strongly near the band center in the case of antisymmetric coupling and near the band edges ($E = \pm 2t^{(0)}$) in the case of symmetric coupling, as was shown in Ref. 23.

To estimate the optical processes associated with the optical conductivity peak, we calculated the mean energy of the involved carriers according to $\langle E_{\mathrm{ini}} \rangle = \frac{\int \Xi(E, E-\hbar\omega_1) E dE}{\int \Xi(E, E-\hbar\omega_1) dE}$ and $\langle E_{\mathrm{fin}} \rangle = \langle E_{\mathrm{ini}} \rangle - \hbar\omega_1$. The results for the case where we consider the e-ph coupling symmetry relevant for the rubrene crystal, $c = 0.44$, are displayed in Fig. 5(b). When the Fermi level falls into the band region ($E_{\mathrm{f}} < 2t^{(0)}$), $\langle E_{\mathrm{ini}} \rangle$ is very close in energy to $E_{\mathrm{f}}$ and both $\langle E_{\mathrm{ini}} \rangle$ and $\langle E_{\mathrm{fin}} \rangle$ evolve linearly with $E_{\mathrm{f}}$ as well, indicating that the absorptions related to the optical conductivity peak mainly take place near the Fermi level. $\langle E_{\mathrm{ini}} \rangle$ and $\langle E_{\mathrm{fin}} \rangle$ become independent of $E_{\mathrm{f}}$ when the Fermi level goes beyond the band region ($E_{\mathrm{f}} > 2t^{(0)}$); then, the results in the low carrier concentration regime are recovered. In this context, the optical conductivity peak can be understood in terms of *optical absorptions from the tail states,* a process in which the carriers are optically excited from significantly localized states in the band tail to less localized states within the band, as schematically illustrated in Fig. 5(c). Interestingly, as the absorptions take place around the band tail, the energy of the optical conductivity peak is closely related to the width of the tail. In the present model, the DOS near the band tail can be approximately expressed as:

$$\rho(E) = \rho(E_{\mathrm{edge}}) \exp\left[-(E - E_{\mathrm{edge}})^2 / 2E_{\mathrm{tail}}^2\right], \tag{12}$$

where $E_{\mathrm{edge}} = 2t^{(0)}$ and $E_{\mathrm{tail}}$ denote the band edge and the band-tail width, respectively. In Fig. 6, we plot the optical conductivity peak energy $\hbar\omega_1$ as well as the tail width $E_{\mathrm{tail}}$ as a function of temperature. At 300 K, we calculated that $E_{\text{tail}}$ is about 26 meV, in excellent agreement with the experimental estimate of about 25 meV for the shallow trap states caused by thermal disorders in the rubrene crystal.[45] As both $\hbar\omega_l$ and $E_{\text{tail}}$ evolve linearly with temperature, a linear relation between $\hbar\omega_l$ and $E_{\text{tail}}$ (with $\hbar\omega_l/E_{\text{tail}} = 1.6$) is obtained, as shown in the inset of Fig. 6. Such a behavior could be accessed experimentally via a temperature-dependent measurement in OFET configuration at low-gate voltage.

We note that different theoretical perspectives on optical conductivity and its connection with transport properties in organic crystals can be found in the literature. For instance, through a dynamics approach in which the phonons were also treated classically, Ciuchi *et al.* demonstrated that the finite-energy peak in the optical conductivity represents a clear manifestation of "transient localization" of the charge carriers caused by thermal disorder.[21] By using a similar model, Cataudella and co-workers pointed out that the effects of quantum lattice fluctuations, which can be reflected in the Lorentzian broadening of the *δ*-function in the Kubo formula [see Eq. (4)], should be taken into account when estimating the carrier dynamics in the low-energy regime.[22] A consensus that can be reached from these studies and the present work here, is that thermal disorders, rather than polaronic effects, play a crucial role in determining the optical and transport properties of high-mobility organic semiconductors.

Finally, we stress that, at room temperature, all the quasiparticle states in the rubrene crystal are of an incoherent nature since the optical conductivity peak remains at a finite energy regardless of the position of the Fermi level throughout the electronic spectrum [see Fig. 5(a)]. In fact, the ARPES results for the rubrene and pentacene crystals[6,7] also indicate that the broadening of the quasiparticle states at room temperature is comparable to the total width of the electronic band, which suggests that the lifetime of the corresponding state is too short to support delocalized carriers. This is likely a result of the anisotropic electronic properties of these materials, which in general lead to quasi-one- or two-dimensional charge transport, under which conditions all the electronic states become truly localized in the presence of (static) disorders.[46] However, at very low temperatures where the quantum lattice fluctuations become significant,[22] the finite-energy peak in the optical conductivity may disappear and a Drude-like optical response is expected to recover if the fluctuations are large enough.

## V. CONCLUSIONS AND DISCUSSIONS

We have investigated the infrared optical properties of the rubrene crystal based on a multiscale modeling approach. Our calculations reproduce very well the experimental data that point simultaneously to a nonmonotonic optical conductivity spectrum and small optical effective masses, two ingredients that appear at first sight to be incompatible within the same system. By focusing on e-ph interactions in the static approximation, we uncover an unusual decrease of the optical effective masses with increasing nonlocal e-ph coupling as a result of the broadening (rather than narrowing) of the electronic bandwidths. On the other hand, as the large thermal disorders at room temperature significantly localize the charge carriers and thereby cause a breakdown of band-like transport, the optical absorptions of localized carriers near the band tail are demonstrated to be responsible for the nonmonotonic optical conductivity spectra. These results strongly suggests that the presence of apparently light carriers with small effective mass, as required for achieving high mobility (through $\mu = e\tau/m$), does not necessarily point to band-like transport in organic crystals, contrary to what was previously thought.[6,7,26] In fact, the e-ph couplings have a stronger impact on the relaxation time $\tau$,[23] rather than on the effective mass $m$, making $m$ less significant in determining the transport properties. Our findings are expected to be generally applicable to high-mobility organic semiconductors and should stimulate experimental interest in their verification.

It is important to point out that the definition of the optical effective mass in Eq. (6) is strictly valid only when the Drude model in Eq. (5) is applicable to the system.[27] The optical conductivity lineshapes, however, strongly deviate from a Drude response when the e-ph coupling is significant (see Fig. 1). This underlines that, while the standard Drude model was widely employed to extract the optical effective mass, it may not apply to the rubrene crystal (and similar materials) because of the breakdown of band-like transport; accordingly, the optical effective masses become less well-defined even though very small values of the effective masses can still be obtained. The development of a more adequate theory thereby remains an open problem.

## ACKNOWLEDGMENTS

This work has been supported by the National Science Foundation under Award No. DMR-0819885 of the MRSEC Program. The computational resources have been made available in part by the National Science Foundation under Award No. CHE-0946869 of the CRIF Program. Yuanping Yi acknowledges financial support from the Natural Science Foundation of China (Award No. 21373229).

**Table I.** Parameter values estimated for the rubrene crystal.

| $t^{(0)}$ (meV) | $L$ (meV) | $c$ | $\hbar\omega_{ph}$ (cm$^{-1}$) |
|---|---|---|---|
| 95 | 21.4 | 0.44 | 50 |

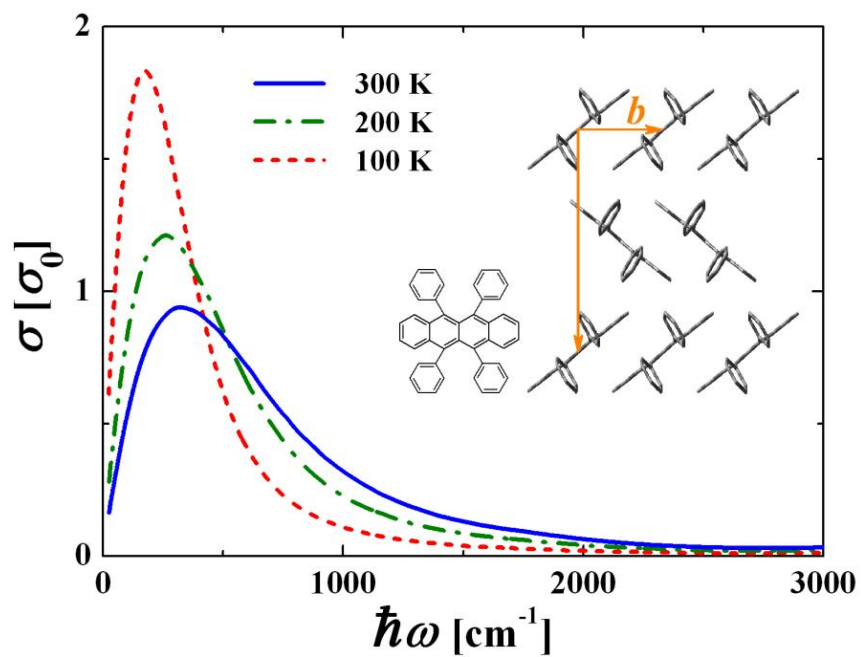

**FIG. 1 (color online).** Optical conductivity spectra along the *b* (stacking) direction of the rubrene crystal calculated at different temperatures. The inset shows the chemical and crystal structures of rubrene. Here, $\sigma_0 = \pi e^2 b / \hbar N$.

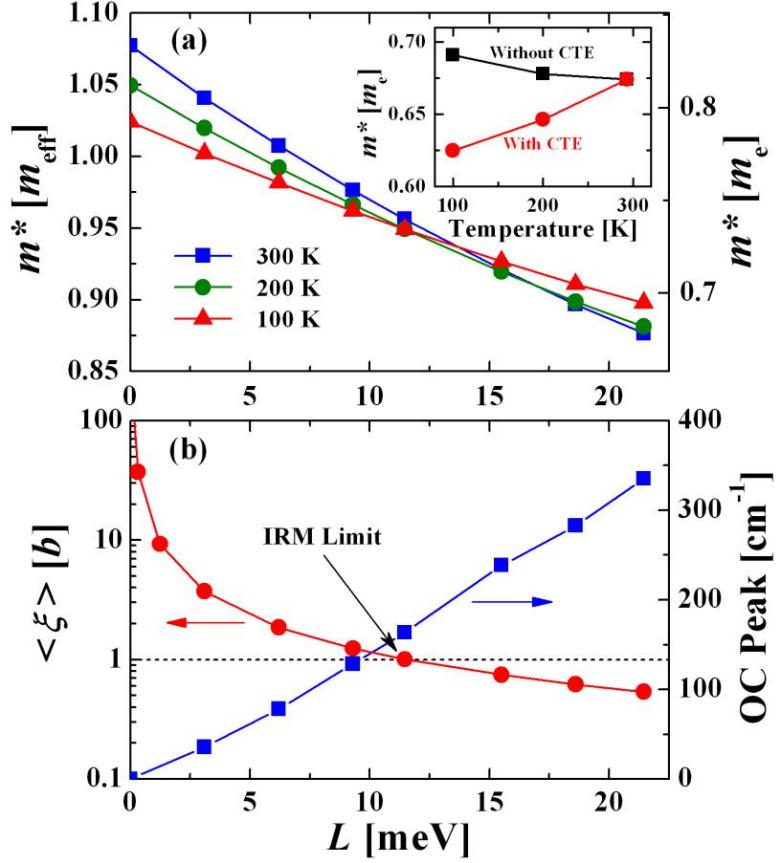

**FIG. 2 (color online).** (a) Optical effective mass at different temperatures and (b) carrier mean free path (circles and left axis) and optical conductivity peak energy (squares and right axis) at 300 K as a function of e-ph coupling strength. The inset in (a) presents the optical effective mass with and without considering the impact of crystal thermal expansion (CTE), as a function of temperature.

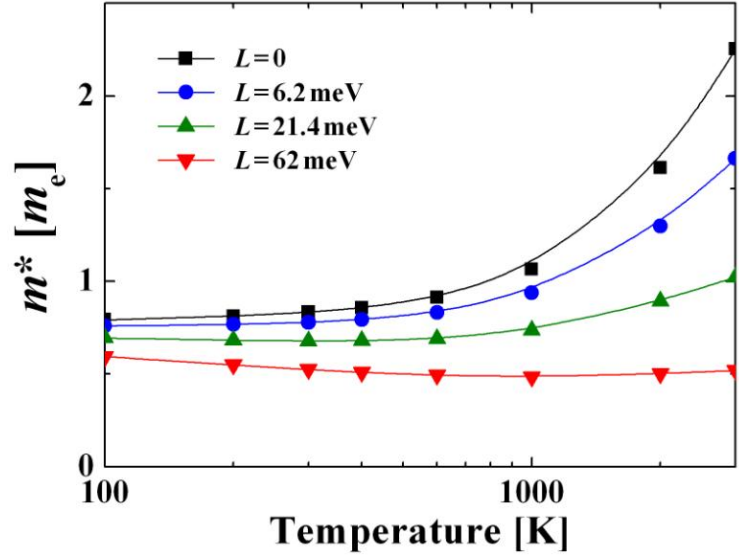

**FIG. 3 (color online).** Optical effective mass obtained from Eq. (8) as a function of temperature for different strengths of e-ph coupling.

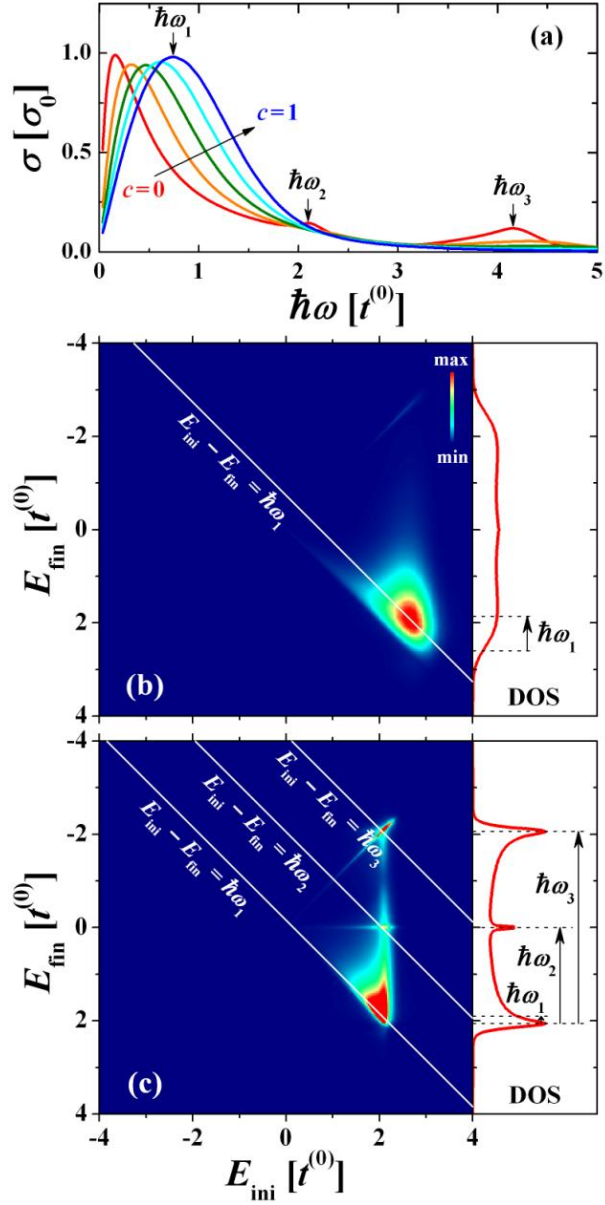

**FIG. 4 (color online).** (a) Optical conductivity spectra as a function of the e-ph coupling symmetry parameter $c$ (in intervals of 0.25); (b) and (c) energy-resolved $\Xi(E_{ini}, E_{fin})$ (left part) and optical absorptions of the optical conductivity peaks projected onto the DOS (right part) for:

(b) symmetric coupling ( $c=1$ ) and (c) antisymmetric coupling ( $c=0$ ). Here, $T = 300$ K; $\sigma_0 = \pi e^2 b / \hbar N$; the white lines in (b) and (c) denote the paths of the integral in Eq. (9) for the optical conductivity peaks.

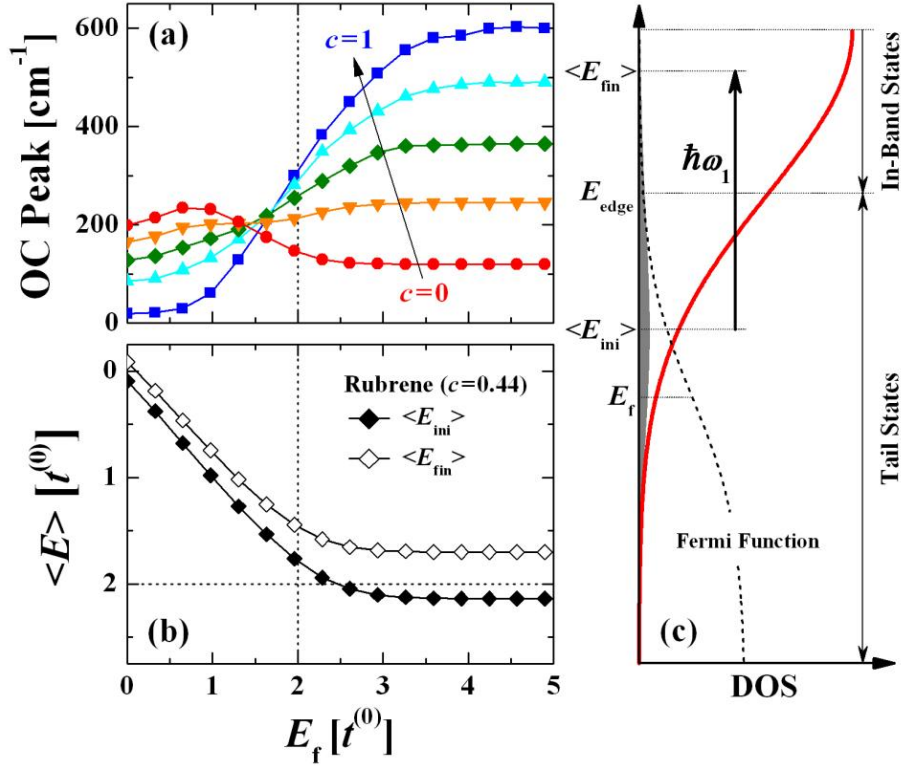

**FIG. 5 (color online).** Fermi-level dependence of (a) optical conductivity peak energy as a function of the e-ph coupling symmetry parameter $c$ (in intervals of 0.25) and (b) mean energy of the initial and final states associated with the absorption of the optical conductivity peak in the rubrene crystal; (c) schematic diagram of the optical absorption near the band tail. Here, $T = 300$ K; in (a) and (b), the dashed lines indicate the band edge in the absence of e-ph coupling; in (c), the filled region represents the occupation of carriers and $E_{\text{edge}}$ denotes the energy of the band edge.

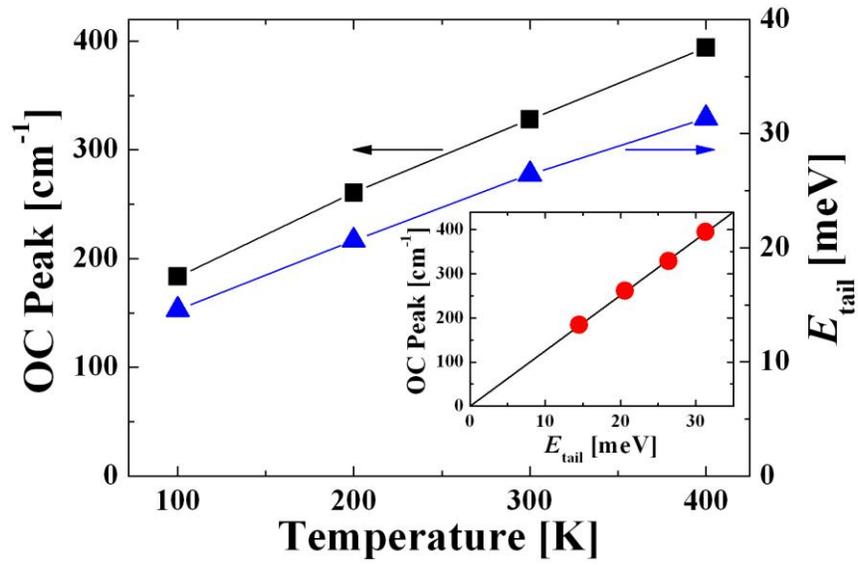

**FIG. 6 (color online).** Temperature dependence of the optical conductivity peak energy (squares and left axis) and the band-tail width (triangles and right axis) in the rubrene crystal in the limit of low carrier concentration. The inset illustrates the linear relationship between the optical conductivity peak energy and the band-tail width.